\documentstyle[12pt,a4wide,amsmath,amssymb,epsf]{article}
\newcommand{\nin}{\noindent}
\newcommand{\be}{\begin{equation}}
\newcommand{\ee}{\end{equation}}
\newcommand{\bea}{\begin{eqnarray}}
\newcommand{\eea}{\end{eqnarray}}
\newcommand{\br}{\hskip .25cm/\hskip -.25cm}
\newcommand{\hf}{\frac{1}{2}}
\newcommand{\nonu}{\nonumber\\}

\newcommand{\ol}{\overline}

\begin{document}

\begin{center}

{\Large{\bf From $N=1$ to $N=2$ supersymmetries\\
in 2+1 dimensions}}\\

\vspace{1cm}

{\bf Jean Alexandre}\\
Physics Department, King's College \\
WC2R 2LS, London, UK\\
jean.alexandre@kcl.ac.uk

\vspace{3cm}

{\bf Abstract}

\end{center}

\vspace{.2cm}

Starting from $N=1$ scalar and vector supermultiplets in 2+1 dimensions, we
construct superfields which constitute Lagrangians invariant under $N=2$
supersymmetries. We first recover the $N=2$ supersymmetric Abelian-Higgs model
and then the $N=2$ pure super Yang-Mills model. The conditions for this 
elevation are consistent with previous results found by other authors.

\vspace{3cm}

$N=2$ supersymmetry in 2+1 dimensions had been studied
\cite{affleck} in order to
investigate the supersymmetrization of the instantons effects which lead
to a linear confinement \cite{poliakov}.
Since then, systematic studies of $N=2$ supersymmetry in 2+1
dimensions have been done
in \cite{strassler} where exact results were obtained,
such as
superpotentials and topologies of moduli spaces in various cases.

These exact results can be derived
since $N=2$ supersymmetry in 2+1 dimensions can be obtained by
dimensional reduction of $N=1$ supersymmetry in 3+1 dimensions \cite{siegel}
and thus has similar properties, such as non-renormalization theorems.
These theorems are not
present for $N=1$ supersymmetry in 2+1 dimensions
and it is thus interesting to study its elevation to $N=2$.
In this context,
it was shown that the presence of topologically
conserved currents
leads to a centrally extended $N=2$ superymmetry, the
central charge of the
superalebra being the topological charge \cite{hlousek}.
In \cite{edelstein}, the $N=1$ supersymmetric Abelian-Higgs model
was considered and it was shown that the on-shell Lagrangian can be
extended to the $N=2$
Abelian-Higgs model if a relation is imposed between the gauge
coupling
and the Higgs self-coupling. Such a condition is in general expected in a
$N=2$ invariant theory built out of $N=1$ Lagrangians \cite{sohnius}.
Another example of extension 
was given in \cite{ams2} where $N=1$ supersymmetries of composite
operators
was elevated to a $N=2$ Abelian model, up to irrelevant operators.
In this work, the coupling of matter to the gauge field
was obtained with higher order composites, simulating the dynamical
generation of the $N=2$
supersymmetry that would occur after an appropriate functional
integration of a
gauge field coupled to the original $N=1$ supermultiplets.

We propose here another illustration of the supersymmetry extension,
with the superfield construction of $N=2$ Lagrangians in
terms of $N=1$
scalar and vector superfields.
The $N=1$ superspace in 2+1 dimensions contains only one real
two-component Grassmann coordinate $\theta$
and the invariant actions are integrals over superspace which involve $\int
d^2\theta$.
The Lagrangians are constructed out of superfields which mix the
original $N=1$
superfields in such a way that a $N=1$ supersymmetric transformation
on the original superfields
leaves the $N=2$ Lagrangians invariant. 

We first consider the pure $U(1)$ case and
then add matter so as to construct the
Abelian Higgs model, using a Fayet-Iliopoulos term.
We finally make a superfield construction for the
$N=2$ pure Yang-Mills Lagrangian.
As will be seen, the present contruction exhibits naturally the
conditions found in \cite{hlousek} and \cite{edelstein} for the elevation of a $N=1$
to a $N=2$ supersymmetry.

\vspace{.5cm}

The gamma matrices are given by
$\gamma^0=\sigma^2,\gamma^1=i\sigma^1,\gamma^2=i\sigma^3$,
where $\sigma^1,\sigma^2,\sigma^3$ are the Pauli matrices, such that
$g^{\mu\nu}=diag(1,-1,-1)$
and $[\gamma^\mu,\gamma^\nu]=-2i\epsilon^{\mu\nu\rho}\gamma_\rho$.
We have the following usual properties, valid for any
2-component complex spinors $\eta,\zeta$:

\be
\eta\zeta=\eta^\alpha\zeta_\alpha=\zeta\eta~~~~\mbox{and}~~~~
\eta\gamma^\mu\zeta=-\zeta\gamma^\mu\eta,
\ee

\nin The 2-component real spinor $\theta$, Grassmann coordinate in the superspace,
satisfies the properties

\bea\label{prop}
&&(\theta\eta)(\theta\zeta)=-\hf\theta^2(\eta\zeta)\nonu
&&\theta\gamma^\mu\theta=0\nonu
&&\theta\gamma^\mu\gamma^\nu\theta=-\theta^2g^{\mu\nu}\nonu
&&\theta\gamma^\mu\gamma^\nu\gamma^\rho\theta=
-i\theta^2\epsilon^{\mu\nu\rho}\nonu
&&\theta\gamma^\mu\gamma^\nu\gamma^\rho\gamma^\sigma\theta=
-\theta^2\left(g^{\mu\nu}g^{\rho\sigma}+g^{\mu\rho}g^{\nu\sigma}
-g^{\mu\sigma}g^{\nu\rho}\right),\nonu
&&\int d^2\theta~\theta^2=1
\eea

In 2+1 dimensions, the $N=1$ scalar superfield and the $N=1$ vector
superfield in the Wess-Zumino gauge
are respectively given by

\bea
\Phi&=&\rho+(\theta\xi)+\hf\theta^2 D\nonu
V_\alpha&=&i(\br A\theta)_\alpha+\hf\theta^2\chi_\alpha,
\eea

\nin where all the fields are real.
To form an $N=2$ supermultiplet, we define the complex gaugino
$\lambda=\xi+i\chi$.
The two fermionic degrees of freedom then
balance the two bosonic ones, since $A_\mu$ has one degree of freedom
\cite{binegar}.
The (complex) scalar superfield $G$ containing these degrees of freedom is

\bea
G&=&\Phi+iD^\alpha V_\alpha\\
&=&\rho+(\theta\lambda)+\hf\theta^2 D^2+i\partial_\mu A_\nu
(\theta\gamma^\mu\gamma^\nu\theta)\nonumber,
\eea

\nin where the superderivative is
$D_\alpha=\partial_\alpha+i(\br\partial\theta)_\alpha$ \cite{hitchin}.
We will see that the elevation of the supersymmetry is possible under a gauge 
condition which affects the superfield $G$ and the relevant fundamental
superfield is actually:

\be\label{dbetag}
D^\beta G=-\lambda^\beta-D\theta^\beta-i\partial_\mu
A_\nu(\gamma^\nu\gamma^\mu\theta)^\beta
+i(\br\partial\rho\theta)^\beta+i(\theta\partial_\mu\lambda)
(\gamma^\mu\theta)^\beta.
\ee

\nin With the properties (\ref{prop}), it is easy to see that

\bea
\int d^2\theta D^\beta G D_\beta \ol G&=&
-\hf F^{\mu\nu}F_{\mu\nu}+i\ol\lambda\br\partial\lambda
+\partial_\mu\rho\partial^\mu\rho+D^2\nonu
&&+\left(\partial^\mu A_\mu\right)^2+\mbox{surface term},
\eea

\nin where the surface term is
$\partial_\mu(\ol\lambda\gamma^\mu\lambda)$ and $F_{\mu\nu}=\partial_\mu A_\nu-\partial_\nu A_\mu$.
If the gauge condition $\partial^\mu A_\mu=0$ is imposed, we find then
the $N=2$ Abelian gauge
kinetic term. This gauge condition was found in \cite{hlousek} where
the authors explain that they need to choose a
gauge in which the vector superfield satisfies $D^\alpha V_\alpha=0$ so as to 
construct a superfield containing a topological current and two supercurrents
which are at the origin of the $N=2$ structure. This condition implies then for the gauge field
component that $\partial^\mu A_\mu=0$, where $A_\mu$ is given the role of the
topologically conserved current. It is then natural that we find here the same condition, 
which should be independent of the dynamics. Indeed, it was explicitely shown for the
$CP^1$ model in \cite{hlousek} and will be found again for the non-Abelian dynamics in the 
present article. Note here that, with the condition $D^\alpha V_\alpha=0$,
the superfield $G$ reduces to $\Phi$, showing that the fundamental superfield is actually $D^\beta G$,
which is not affected by this gauge condition, as can be seen with Eq.(\ref{dbetag}) \footnote{I thank the 
referee for pointing this out.}.
  
Disregarding the surface term, the expected $N=2$ Lagrangian is then
expressed in terms of the original $N=1$ superfields as follows:

\bea\label{lgauge}
{\cal L}_{gauge}&=&\hf\int d^2\theta D^\beta G D_\beta \ol G\nonu
&=&-\frac{1}{4}F^{\mu\nu}F_{\mu\nu}+\frac{i}{2}\ol\lambda\br\partial\lambda
+\hf \partial_\mu\rho\partial^\mu\rho+\hf D^2.
\eea

\vspace{.5cm}

Matter is included with a complex $N=1$ scalar superfield $Q$:

\be
Q=\phi+(\theta\psi)+\hf\theta^2 F.
\ee

\nin So as to avoid the generation
of parity violating terms in the quantum corrections,
we can introduce an even number of superfields \cite{strassler}, but we do not consider
this problem here. The interested reader can find a review of supersymmetric Chern-Simons 
theories in \cite{nishino}.
We remind that a $N=1$ scalar superfield in 2+1 dimensions cannot be
chiral:
since $\theta$ is real, the chirality
condition $D^\alpha Q=0$ would constraint the space-time dependence of
the component fields
$\phi,\psi,F$ \cite{hitchin}.

The derivatives of the fields are obtained with the highest component
of
$D^\alpha Q D_\alpha \ol Q$ which reads

\be
\left.D^\alpha Q D_\alpha \ol Q\right|_{\theta^2}=
\theta^2\left(\partial_\mu\phi\partial^\mu\phi+i\ol\psi\br\partial
\psi+F\ol F+
\mbox{surface term}\right),
\ee

\nin where the surface term is
$\partial_\mu(\ol\psi\partial^\mu\psi)$.
The coupling to the gauge multiplet
is obtained with the highest components of the following superfields:

\bea
\left.D^\alpha QV_\alpha\ol Q\right|_{\theta^2}&=&-\hf\theta^2
\left(\ol\phi(\psi\chi)-\ol\phi A_\mu\partial^\mu\phi+i\ol\psi\br
A\psi\right)\nonu
\left.Q\Phi\ol Q\right|_{\theta^2}&=&\hf\theta^2
\left(\phi\ol\phi D-\phi\ol\psi\xi-\ol\phi\psi\xi+\rho\phi\ol
F+\rho\ol\phi F-
\rho\ol\psi\psi\right)\nonu
\left.QV^\alpha V_\alpha\ol Q\right|_{\theta^2}&=&-\theta^2\phi\ol\phi
A^\mu A_\mu,
\eea

\nin such that the matter Lagrangian is

\bea\label{lmat}
{\cal L}_{matter}
&=&\hf\int d^2\theta\left\{(D^\alpha-igV^\alpha)Q(D_\alpha+igV_\alpha)\ol Q
+2gQ\Phi\ol Q\right\}\nonu
&=&\frac{i}{2}\ol\psi\br D\psi+\hf D_\mu\phi
D^\mu\ol\phi-\frac{g}{2}(\phi\ol\psi\ol\lambda+
\ol\phi\psi\lambda)-\frac{g}{2}\rho\ol\psi\psi\nonu
&&+\frac{g}{2}\phi\ol\phi D+\hf F\ol F+\frac{g}{2}\rho\phi\ol F+\frac{g}{2}\rho\ol\phi F,
\eea

\nin where $g$ is a dimensionfull gauge coupling and
$D_\mu=\partial_\mu+ig A_\mu$. The Lagrangian 
(\ref{lmat}) was found in \cite{siegel} as
a consequence of the dimensional reduction of a $N=1$ theory in 3+1 dimensions.
It was also derived in \cite{ivanov} where the $N=2$ Lagrangian
is expressed with $N=1$ superfields. In both these works, the authors start from $N=2$, 
and do not elevate an initial $N=1$ Lagrangian to $N=2$; hence they do not find
any constraint. The reader can find in \cite{gates} a discussion of the relation between $N=1$, $N=2$ and $N=4$
supersymmetries in 1+1, 2+1 and 3+1 dimensions.

\vspace{.5cm}

We can recover the
scalar interactions if we write the Lagrangians (\ref{lgauge}) and
(\ref{lmat}) on-shell.
We write for this the equations of motion of the auxiliary fields $D$
and $F$:

\bea
\ol F+g\rho\ol\phi&=&0\nonu
D+\frac{g}{2}\phi\ol\phi&=&0,
\eea

\nin such that the terms depending on the auxiliary fields
lead to the following potential

\bea
\left({\cal L}_{gauge}+{\cal L}_{matter}\right)_{pot}&=&
\hf D^2+\frac{g}{2}\phi\ol\phi D+\hf F\ol F+\frac{g}{2}\rho\ol\phi F+
\frac{g}{2}\rho\phi\ol F\nonu
&=&-\frac{g^2}{2}\rho^2\phi\ol\phi-\frac{g^2}{8}(\phi\ol\phi)^2.
\eea

The Abelian Higgs model is obtained by adding a
Fayet-Iliopoulos term which in
the present context is

\be
{\cal L}_{F.I.}=-\frac{g}{2}\phi_0^2\int d^2\theta(G+\ol G)
=-g\phi_0^2\int d^2\theta \Phi=-\frac{g}{2}\phi_0^2 D,
\ee

\nin where $\phi_0$ is a real parameter.
The addition of this term to the Lagrangian leads to the following
equation of motion
for the auxiliary field $D$:

\be
D+\frac{g}{2}\phi\ol\phi-\frac{g}{2}\phi_0^2=0,
\ee

\nin such that we obtain the expected gauge-symmetry breaking potential

\be\label{lah}
\left({\cal L}_{gauge}+{\cal L}_{matter}+{\cal L}_{F.I.}\right)_{pot}
=-\frac{g^2}{2}\rho^2\phi\ol\phi-\frac{g^2}{8}\left(\phi\ol\phi-\phi_0^2\right)^2.
\ee

\nin Note that the Higgs self-coupling is $g^2/8$, what was found in \cite{edelstein}
as a consistency condition for the elevation of the $N=1$ on-shell Lagrangian
to $N=2$.
The result (\ref{lah}) shows that the moduli space contains a
Higgs branch only,
where the vacuum expectation values of the scalar fields satisfy

\be
<\phi\ol\phi>=\phi_0^2 ~~~~\mbox{and}~~~~ <\rho>=0.
\ee

\vspace{.5cm}

The extension to a non-Abelian gauge group necessitates the
introduction of quadratic
superfields to generate the interactions. We will consider $SU(N)$
dynamics, with
structure constants $f^{abc}$ and coupling constant $g$.
A non-Abelian supermultiplet contains gauginos and scalars in the
adjoint representation, so that
the starting point is the set of scalar and vector $N=1$ superfields

\bea
\Phi^a&=&\rho^a+(\theta\xi^a)+\hf\theta^2D^a\nonu
V_\alpha^a&=&i(\br A^a\theta)_\alpha+\hf\theta^2\chi^a_\alpha,
\eea

\nin where $a=1,...,N^2-1$ is the gauge indice.
We then introduce the complex superfields

\be
G^a=\Phi^a+iD^\alpha V^a_\alpha,
\ee

\nin and, as in the Abelian case, the derivatives of the component fields are obtained with
the term $D^\beta G^aD_\beta\ol G^a$, provided that the gauge condition
$\partial^\mu A_\mu^a=0$ holds, which shows again that the fundamental superfield is 
actually $D^\beta G^a$ and not $G^a$.
To generate the interactions of the superpartners, we will add to
$D^\beta G^a$ linear combinations of the following two superfields

\be
G^bV^{c\beta},~~~~~~~~D^\beta(V^{b\alpha} V^c_\alpha),
\ee

\nin and the remaining terms for the covariant derivatives
are obtained with the products

\bea\label{quad1}
\left.f^{abc}D^\beta \ol G^a G^b V^c_\beta\right|_{\theta^2}&=&
\theta^2f^{abc}\left(-\rho^b\xi^a\chi^c+\hf(i\ol\lambda^b\br
A^c\lambda^a+\mbox{c.c}.)
-2\rho^b\partial^\mu\rho^a A_\mu^c\right)\nonu
\left.f^{abc}f^{ade}G^bV^{c\beta}\ol G^d V^e_\beta\right|_{\theta^2}&=&
f^{abc}f^{ade}\theta^2\rho^b\rho^d A_\mu^c A^{e\mu}.
\eea

\nin The term (\ref{quad1}) also generates the Yukawa interactions
since

\be
2\rho^b\xi^a\chi^c=\rho^b(i\ol\lambda^a\lambda^c+\mbox{c.c.}).
\ee

\nin The non-Abelian gauge kinetic term is obtained with the products

\bea
\left.f^{abc}D^\beta\ol G^a
D_\beta(V^{b\alpha}V^c_\alpha)\right|_{\theta^2}&=&
2f^{abc}(\partial_\mu A_\nu^a) A_\rho^b
A_\sigma^c(\theta\gamma^\mu\gamma^\nu\gamma^\rho\gamma^\sigma\theta)\nonu
\left.f^{abc}f^{ade}D^\beta(V^{b\alpha}
V^c_\alpha)D_\beta(V^{b\alpha}V^c_\alpha)\right|_{\theta^2}&=&
f^{abc}f^{ade}A_\mu^b A_\nu^c A_\rho^d A_\sigma^e
(\theta\gamma^\mu\gamma^\nu\gamma^\rho\gamma^\sigma\theta),
\eea

\nin since we have, using the properties (\ref{prop}) and
$f^{abc}+f^{acb}=0$,

\bea\label{eq25}
&&f^{abc}\int d^2\theta(\partial_\mu A_\nu^a) A_\rho^b A_\sigma^c
(\theta\gamma^\mu\gamma^\nu\gamma^\rho\gamma^\sigma\theta)\nonu
&=&f^{abc}(\partial^\mu A_\mu^a) A_\nu^b A^{c\nu}+
f^{abc}A_\mu^b A_\nu^c(\partial^\mu A^{a\nu}-\partial^\nu A^{a\mu}),
\eea

\nin and

\bea
&&f^{abc}f^{ade}\int d^2\theta A_\mu^b A_\nu^c A_\rho^d A_\sigma^e
(\theta\gamma^\mu\gamma^\nu\gamma^\rho\gamma^\sigma\theta)\nonu
&=&2f^{abc}f^{ade}A_\mu^b A^{d\mu}A_\nu^c A^{e\nu}.
\eea

\nin With the gauge condition $\partial^\mu A_\mu^a=0$, 
the first term in the right-hand side of Eq.(\ref{eq25}) vanishes and only the
expected term remains.
Gathering these results, we find that the extension to an
$N=2$ pure
super-Yang-Mills Lagrangian is given by

\bea
{\cal L}_{Y.M.}&=&\hf\int d^2\theta\left|D^\beta
G^a+gf^{abc}\left(G^bV^{c\beta}
+\frac{i}{2}D^\beta(V^{b\alpha}V_\alpha^c)\right)\right|^2\nonu
&=&-\frac{1}{4}F^{a\mu\nu}F_{\mu\nu}^a+\frac{i}{2}\ol\lambda^a\br D\lambda^a
+\hf D^\mu\rho^a D_\mu\rho^a\nonu
&&-\frac{g}{2}f^{abc}\left(i\rho^b\ol\lambda^a\lambda^c+\mbox{c.c.}\right)+\hf D^aD^a,
\eea

\nin where $F^a_{\mu\nu}=\partial_\mu A_\nu^a-\partial_\nu A_\mu^a+gf^{abc}A_\mu^b A_\nu^c$
and $D_\mu(...)^a=\partial_\mu(...)^a+gf^{abc}A_\mu^b(...)^c$.

\vspace{1cm}

To conclude, let us stress the central point of these results. Whereas 
the elevation of a $N=1$ to a $N=2$ supersymmetry was shown explicitely for 
the $CP^1$ model in \cite{hlousek} and for the Abelian Higgs model in \cite{edelstein},
we do not start here with any specific dynamics but instead build directly $N=2$
off-shell Lagrangians with $N=1$ superfields.
This allows us to generate different dynamics and we generalize the elevation 
to a $N=2$ non-Abelian theory. Clearly, one could consider with the same method
other $N=2$ dynamics.

Finally, this work might be used in the context of effective models
for high-temperature (planar) superconductivity \cite{mavromatos},
where the initial $N=1$ supermultiplets are built out of composites of spinons and holons in the 
spin-charge separation framework.

\vspace{1cm}

\nin {\bf Acknowledgments} This work is supported by the Leverhulme Trust (U.K.) and I
would like to thank Sarben Sarkar and Nick Mavromatos for introducing me to this
subject.

\end{document}